\documentclass[epj]{webofc}
\usepackage[varg]{txfonts}   % Web of Conferences font
\wocname{EPJ Web of Conferences}
\woctitle{Resonance Workshop at Catania}
\begin{document}
 \title{Thermal model for Pb+Pb collisions at $\sqrt{s}_{NN} = 2.76$~TeV with explicit treatment of hadronic ground states
%
% The effects of adding the ground state for bosons to the thermal model in the description of Pb+Pb collisions at $\sqrt{s}_{NN} =
% 2.76$~TeV at the LHC
}

%
% Transverse momentum spectra of hadrons produced in Pb+Pb collisions at $\sqrt{s}_{NN} = 2.76$~TeV in the chemical non-equilibrium model}
%
% subtitle is optionnal
%
%%%\subtitle{Do you have a subtitle?\\ If so, write it here}

\author{Viktor Begun\inst{1,2}\fnsep\thanks{\email{viktor.begun@gmail.com}}}

 \institute{Institute of Physics, Jan Kochanowski University, PL-25406~Kielce, Poland
 \and       Bogolyubov Institute for Theoretical Physics, 03680 Kiev, Ukraine}

\abstract{
  Various explanations of the anomalous proton to pion ratio at the LHC are discussed.
  The special emphasis is set on the Cracow thermal model with single freeze-out.
  This model allows to get a good agreement for both the mean hadron multiplicities and the spectra.
  Moreover, the values of the fit parameters indicate the possibility of pion Bose condensation in
  the most central collisions at the LHC.
  Therefore,  a modification of the thermal framework is proposed that explicitly allows for the condensation
  in the ground state.
  The generalised model makes a link between equilibrium and non-equilibrium thermal models.
  It also suggests that the pion condensation may be formed in the central collisions.
   }
\maketitle
\section{Introduction}
\label{intro} Statistical models are used as the standard tools
for the analysis of heavy-ion and elementary ($e^+e^-$,
$p\bar{p}$, etc.) collisions. These models give a very good
description of mean multiplicities of many hadron species using
only few parameters, for example,
see~\cite{Cleymans:1992zc,Cleymans:1999st,BraunMunzinger:2003zd,Becattini:2003wp,Becattini:1995if,Becattini:1997rv}.
Therefore, it is quite surprising that the new data from the LHC
do not agree with the thermal model prediction for proton
abundances~\cite{Stachel:2013zma}.
Among possible explanations of this problem there are: hadronic
re-scattering effects in the final stage~\cite{Becattini:2012xb},
incomplete list of
hadrons~\cite{Noronha-Hostler:2014usa,Noronha-Hostler:2014aia},
flavor hierarchy at freeze-out~\cite{Chatterjee:2013yga}, and the
non-equilibrium
hadronization~\cite{Petran:2013qla,Petran:2013lja}, see also
\cite{Floris:2014pta}. Herein, we will focus on the latter
explanation, because, as we have shown before
in~\cite{Begun:2013nga,Begun:2014rsa}, it offers a plausible
description of the transverse momentum spectra of the produced
hadrons.

Surprisingly, hydrodynamic models have problems to reproduce the
pion spectra at the LHC as well. The low-$p_T$ pion spectra show
enhancement by about $25\%-50\%$ with respect to the predictions
of different hydrodynamic models,  see the compilation shown by
ALICE in Refs.~\cite{Abelev:2012wca,Abelev:2013vea}. One can
notice that the pions and protons are anti-correlated. If a model
explains protons, it typically underestimates pions. On the other
hand, if a model explains pions, then it overestimates protons.
More recent papers also illustrate this
issue~\cite{Gale:2012rq,Molnar:2014zha}. Only in
Ref.~\cite{vanderSchee:2013pia} the pions are described in the
satisfactory way, however, no results for the protons are given in
this work.

\section{Cracow single-freeze out model}
\label{sec-1}
The Cracow single-freeze out
model~\cite{Broniowski:2001we,Broniowski:2001uk,Rybczynski:2012ed}
allows to solve the problem with the proton/pion ratio and the
problem with the pion spectrum~\cite{Begun:2013nga,Begun:2014rsa}.
The model includes all well established resonances from the PDG.
The masses of resonances and their decays are implemented in the
THERMINATOR Monte-Carlo
code~\cite{Kisiel:2005hn,Chojnacki:2011hb}. The primordial
distribution in the local rest frame has the form:
\begin{equation}\label{fprim}
 f_i ~=~ g_i \int \frac{d^3p}{(2\pi)^3}\frac{1}{\Upsilon_i^{-1}\exp\left(\sqrt{m_i^2+p^2}/T\right)\pm1}~,
\end{equation}
 where $g_i=2s_i+1$ is the degeneracy connected with the spin $s_i$ of the $i$th particle, $p$ is the particle momentum, $m_i$ - mass, and $T$ is the system temperature. The factor $\Upsilon_i$ is expressed by the numbers of light quarks, $N^i_q$, antiquarks, $N^i_{\bar q}$, strange quarks, $N^i_s$, strange antiquarks $N^i_{\bar s}$; baryon and strange charges of  the particle - $B_i$, $S_i$, and the corresponding chemical potentials, $\mu_B$ and $\mu_S$:
\begin{equation}
 \Upsilon_i ~=~ \gamma_q^{N^i_q+N^i_{\bar q}} \gamma_s^{N^i_s+N^i_{\bar s}}  \exp \left( \frac{ \mu_B B_i  + \mu_S S_i}{T}\right)~.
\end{equation}
At the LHC the chemical potentials $\mu_B$ and $\mu_S$ are so
small that one can set them zero. However, the introduction of the
parameters $\gamma_q$ and $\gamma_s$ is equivalent to the
appearance of the non-equilibrium chemical potentials
$\mu_i/T=\ln\gamma_i$:
\begin{equation}\label{upsiNeq2}
 \Upsilon_i ~\simeq~ \gamma_q^{N^i_q+N^i_{\bar q}} \gamma_s^{N^i_s+N^i_{\bar s}}
 ~=~ \exp\left(\frac{\mu_q\left(N_q^i+N_{\bar{q}}^i\right)+\mu_s\left(N_s^i+N_{\bar{s}}^i\right)}{T}\right)~.
%\nonumber
\end{equation}
They are connected with the conservation of the {\it sum} of the
number of quarks and antiquarks during the hadronization process.
Similarly, the usual baryon and strange chemical potentials
$\mu_B$ and $\mu_S$ are connected with the conservation of the
{\it difference} of the quark and antiquark numbers. Such an
effective quark number conservation may appear due to rapid
cooling and hadronization of the fireball. Then the system has no
time to equilibrate and the numbers of quarks and antiquarks are
larger than the equilibrium values.

We note that, the non-equilibrium model may account for
hypothetical heavy particles that decay into multi-pion
states~\cite{Noronha-Hostler:2014usa,Noronha-Hostler:2014aia}.
Equation~(\ref{upsiNeq2}) may describe the equilibrium $p+\bar{p}$
annihilation into 3 pions. One can also notice that the
$\Upsilon_i$ factor is different for each particle. Some particles
are enhanced, while the other are suppressed, compared to the
equilibrium case. Therefore,  Eq.~(\ref{upsiNeq2}) resembles the
modification factors that are obtained in the hadron gas with
rescattering effects~\cite{Becattini:2012xb}.
Equation~(\ref{upsiNeq2}) obviously separates the strange and
non-strange particles. Therefore, it is similar to the model with
two separate freeze-outs for strange and non-strange particles
proposed in Ref.~\cite{Chatterjee:2013yga}.
A QCD mechanism of gluon condensation may also lead to a similar
effect: the creation of low momentum gluons which transform into
pions in the
condensate~\cite{Blaizot:2011xf,Blaizot:2013lga,Gelis:2014tda}.

We consider two physics scenarios: the equilibrium case (EQ),
where $\gamma_q=\gamma_s=1$, and the full non-equilibrium case
(NEQ) with $\gamma_q$ and $\gamma_s$ treated as free parameters.
The spectra are calculated from the Cooper-Frye formula with the special freeze-out hypersurface:
\begin{align}
 \frac{dN}{dy d^2p_T} ~=~ \int d\Sigma_\mu p^\mu f(p \cdot u), &&
 t^2 ~=~ \tau_f^2 + x^2 + y^2 + z^2, && x^2 + y^2 ~\leq~ r^2_{\rm max}~,
%\nonumber
\end{align}
assuming the Hubble-like flow $u^\mu = x^\mu/\tau_f$.

The system volume, temperature, $\gamma_q$, and $\gamma_s$ are
taken from the papers~\cite{Petran:2013qla,Petran:2013lja} and
approximated by the polynomials, for details
see~\cite{Begun:2014rsa}. The combination of the freeze-out time,
$\tau_f$, and the maximum radius squared, $r^2_{\rm max}$, gives
the system volume per unit rapidity, $V=\pi \tau_f r^2_{\rm max}$.
Therefore, the ratio $r_{\rm max}/\tau_f$ is the only one
additional parameter in the model that determines the shape of the
spectra.

\begin{figure}
\centering
 \includegraphics[width=0.48\textwidth,clip]{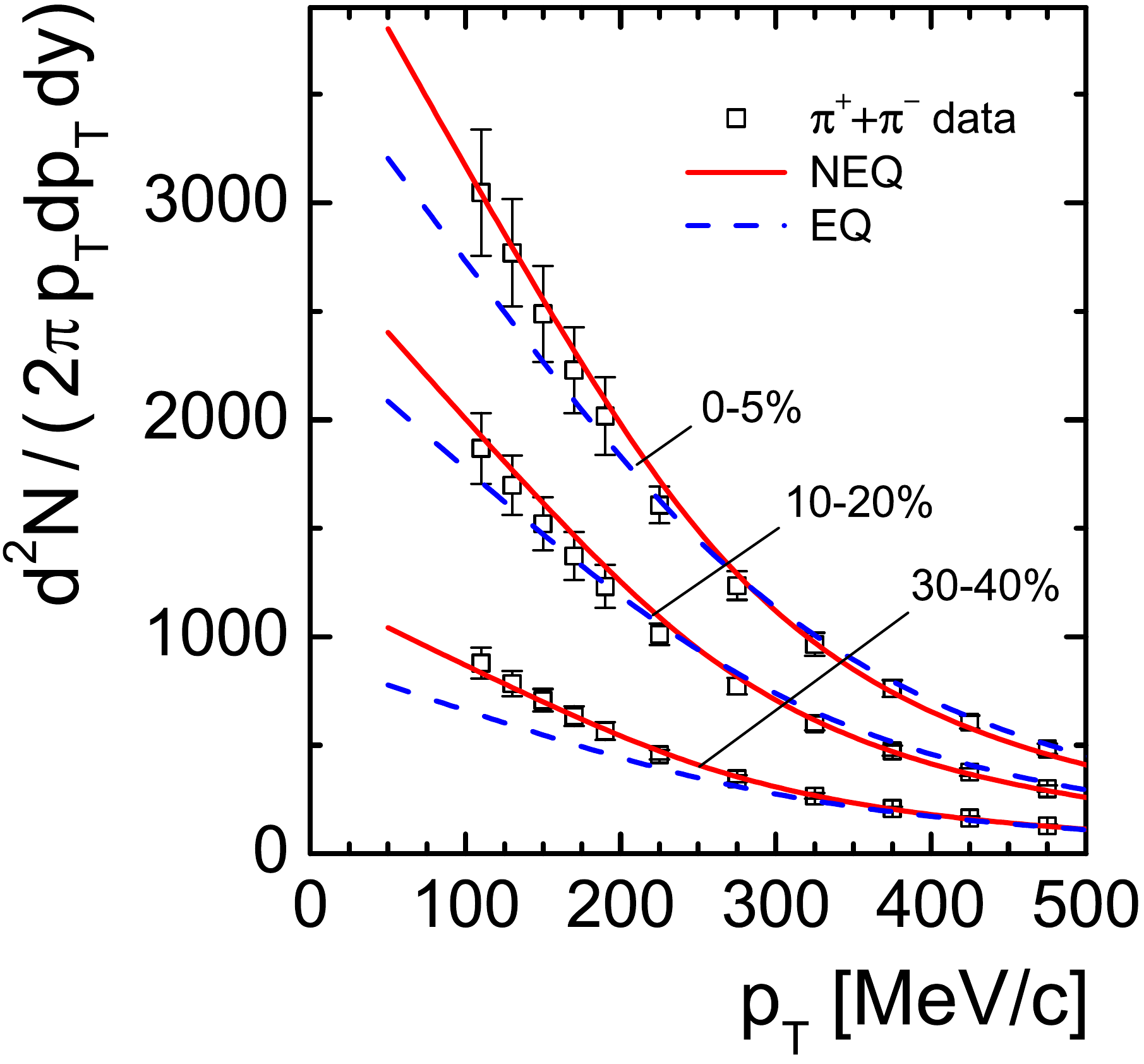}
 \includegraphics[width=0.48\textwidth,clip]{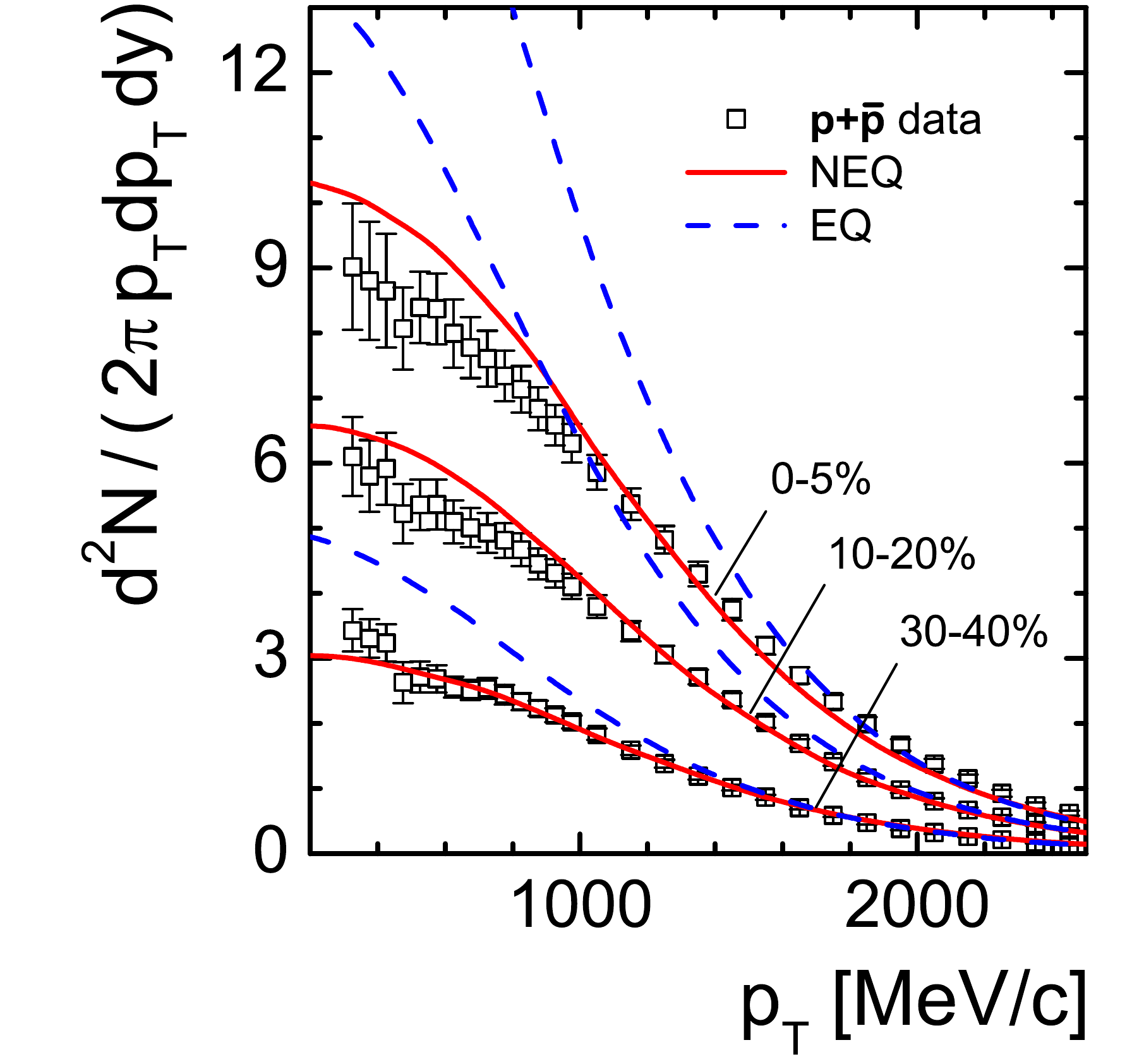}
 \caption{Low $p_T$ spectrum of pions (left) and protons (right) in three centrality windows: $0-5\%$, $10-20\%$, $30-40\%$.
 The data are from~\cite{Abelev:2012wca}.
 The solid line shows the fit obtained for pion and kaon spectra, without protons, in the chemical non-equilibrium Cracow model.
 The dashed line shows the same fit in the equilibrium Cracow model~\cite{Begun:2013nga,Begun:2014rsa}.} \label{fig-1}
\end{figure}

The Cracow model allows to fit the spectra of pions and kaons with
very good accuracy only in NEQ model, see~\cite{Begun:2013nga} and
also Fig.~\ref{fig-1} left. Surprisingly, the proton spectrum
comes out right without extra fitting as a bonus, see
Fig.~\ref{fig-1} right. The increase in the multiplicity of
primordial pions due to $\gamma_q^2>1$ is compensated by the
decrease of volume and temperature in NEQ, see Fig.~\ref{fig-2}.
However, despite of even larger factor for protons,
$\gamma_q^3>1$, their number is much smaller in NEQ than in EQ. It
happens because of decreased contribution from resonance decays,
due to lower temperature in NEQ. This effect is much stronger for
protons, because they are heavier than pions. The yields are given
by the integrals of the corresponding spectra. Therefore, NEQ
model is also better for proton to pion ratio.

The same fit gives the very good agreement for the spectra of
$K_S^0$, $K^*(892)^0$, $\phi(1020)$ mesons and a satisfactory
agreement for the heavy strange particles from the most central to
very peripheral collisions~\cite{Begun:2014rsa}. As we already
mentioned in the Introduction, the simultaneous fit of the pion
and proton spectra is very difficult, and the difference between
EQ and NEQ models drastically increases at low $p_T$, see
Fig.~\ref{fig-1}.
However, even more surprising fact is that the long living
$\phi(1020)$ and the very short living $K^*(892)^0$ come out right
from the fit done for pions and kaons
only~\cite{Begun:2013nga,Begun:2014rsa}. It is a very strong
argument either for the absence of the long rescattering phase
after the freeze-out or for the effective parametrization of the
re-scattering phase by Eq.~(\ref{upsiNeq2}).

%%%%%%%%%%%%%%%%%%%%%%%%%%%%%%%%%%%%%%%%%%%%%%%%%%%%%%%%%%
%
\section{Pion condensation} \label{sec-2}
%
%%%%%%%%%%%%%%%%%%%%%%%%%%%%%%%%%%%%%%%%%%%%%%%%%%%%%%%%%%
%
 There is an upper bound on $\gamma_q$ and $\gamma_s$ because of Bose-Einstein condensation, when the singularities appear in the Bose-Einstein distributions of primordial pions and kaons (\ref{fprim}).
 For pions, the value of $\gamma_s$ is irrelevant, and we find
\begin{equation}
 \gamma_q^{\rm critic} ~=~ \exp\left(\frac{m_{\pi^0}}{2T}\right). \nonumber
\end{equation}
 The fits to the ratios of hadron abundances yield $\gamma_q$ which is very close to the critical. It is equivalent to the pion chemical potential
 \begin{equation}
 \mu_{\pi}~=~2T\ln\gamma_q\simeq 134~\text{MeV}~,
  \nonumber
 \end{equation}
 which is very close to the $\pi^0$ mass, $m_{\pi^0}\simeq$ 134.98~MeV. It may lead for the condensation of the substantial part of $\pi^0$ mesons.

If the chemical potential approaches the mass of a particle,
$\mu\rightarrow m$, the zero momentum level, $p_0=0$, and other
low lying quantum states become important. Therefore,  one should
consider the summation over the low momentum states explicitly.
One can show that in the thermodynamic limit,
$V\rightarrow\infty$, one may keep only the $p_0=0$ term and start
the integration from zero~\cite{Begun:2008hq}:
\begin{align} \label{Ncond}
 N
 %&~=~ \sum_i \frac{g_i}{\exp\left(\frac{\sqrt{p_i^2+m^2}-\mu}{T}\right)-1}\nonumber
 %\\
 %&~\longrightarrow~
 ~=~ \frac{g}{\exp\left(\frac{m-\mu}{T}\right)-1}
 ~+~  V\int_0^{\infty} \frac{d^3p}{(2\pi)^3}\, \frac{g}{\exp\left(\frac{\sqrt{p^2+m^2}-\mu}{T}\right)-1}
 ~=~ N_{\rm cond} ~+~ N_{\rm norm}~
\end{align}
 where $N_{\rm cond}$ is the number of particles in the Bose condensate and $N_{\rm norm}$ is the number of particles in normal states.
 We have added the condensation term from (\ref{Ncond}) to the latest version of SHARE~\cite{Petran:2013dva}, because it is the model that was used to obtain our input parameters,
 $V$, $T$, $\gamma_q$, $\gamma_s$. The obtained non-equilibrium model with the possibility of Bose condensation we call BEC.

\begin{figure*}
\centering
 \includegraphics[width=0.48\textwidth,clip]{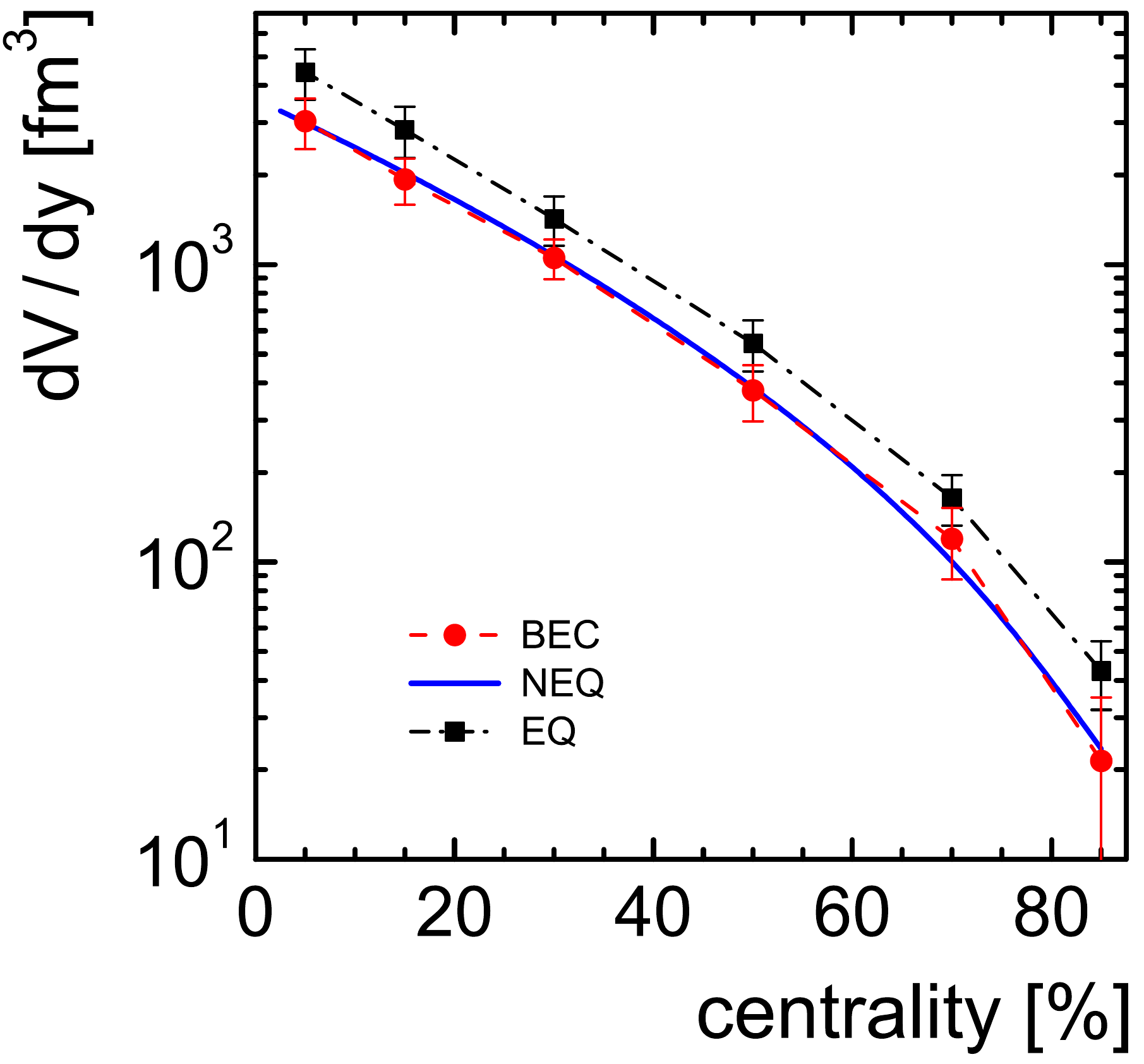}~~
 \includegraphics[width=0.48\textwidth,clip]{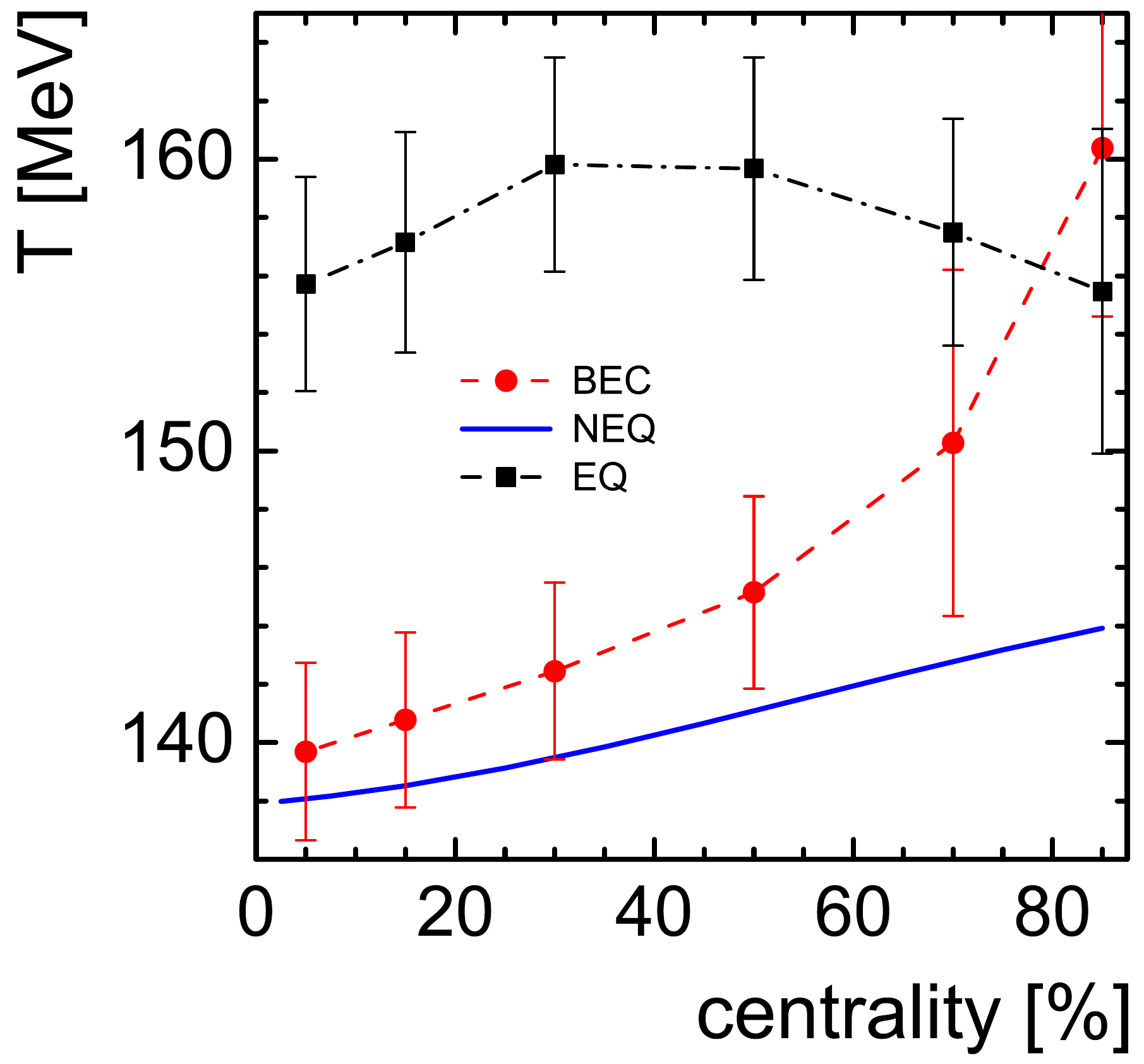}
 \caption{The non-equilibrium parameters from the paper~\cite{Begun:2014rsa}, NEQ, are compared to the new fit in SHARE~\cite{Petran:2013dva} using the equilibrium model, EQ, and the
 non-equilibrium model with the possibility of Bose condnsation, BEC. The left panel shows the system volume, while the right panel shows the system temperature.}
\label{fig-2}
\end{figure*}

\begin{figure*}
\centering
 \includegraphics[width=0.48\textwidth,clip]{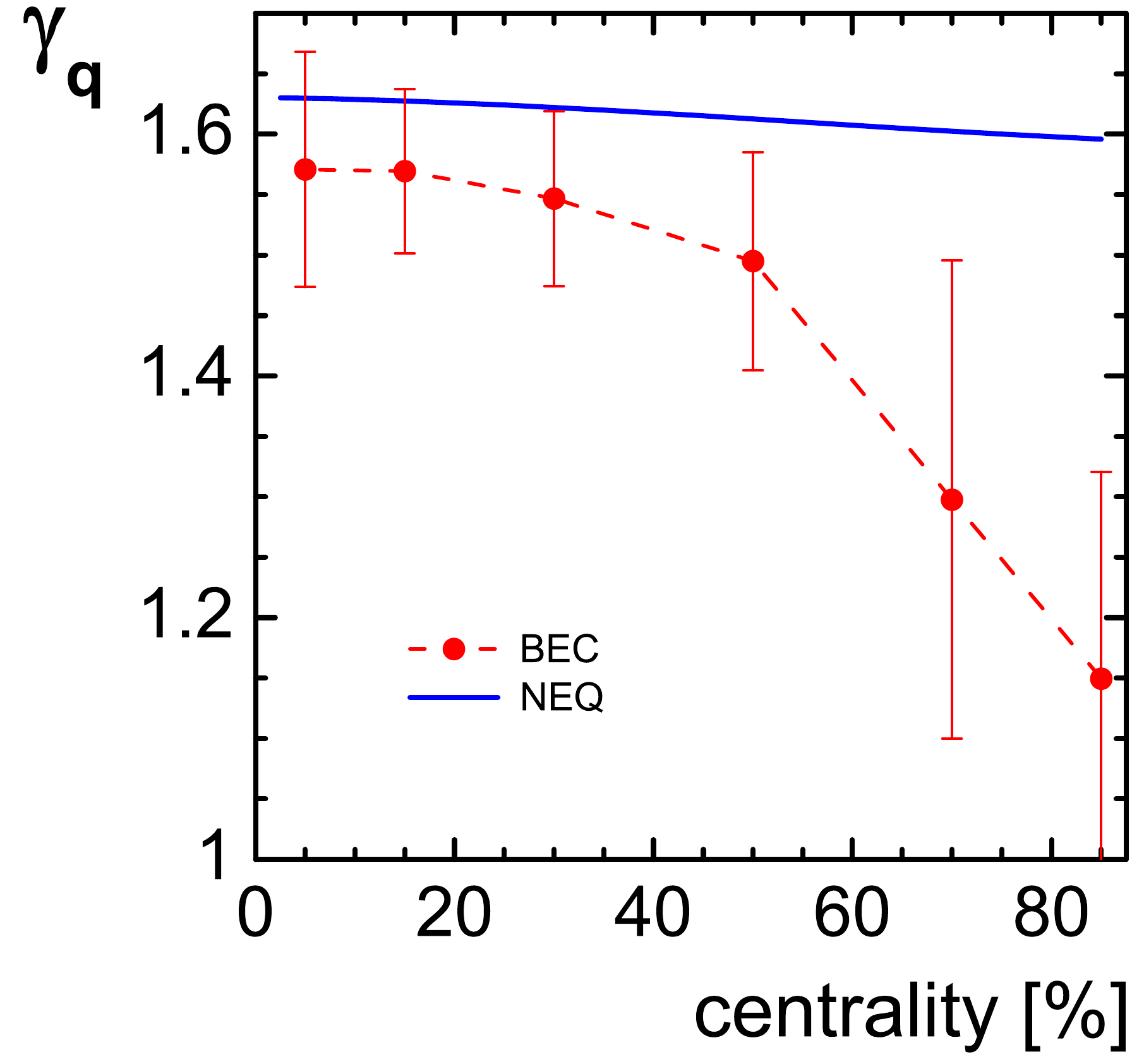}~~
 \includegraphics[width=0.48\textwidth,clip]{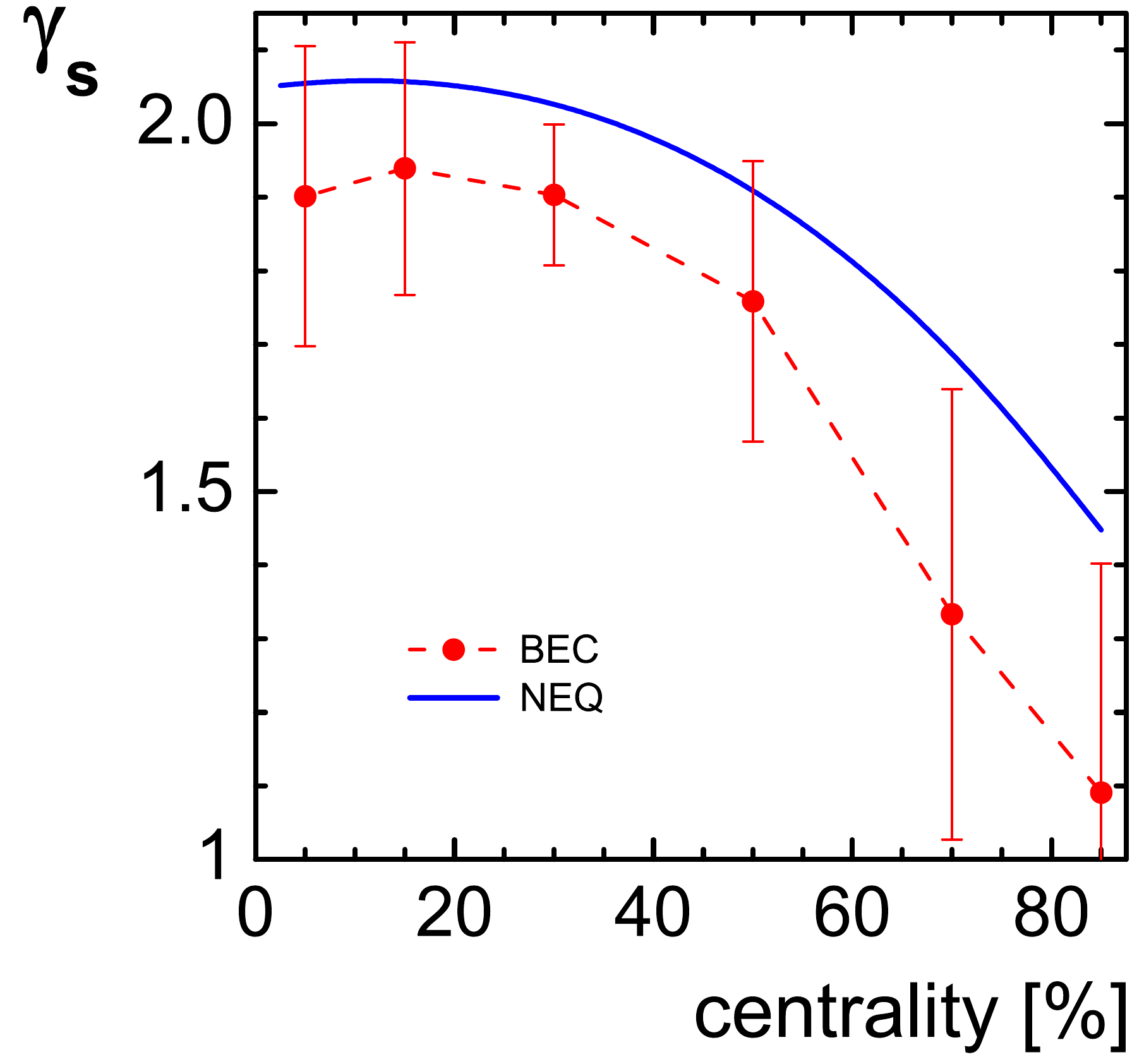}
 \caption{The same as in Fig.~\ref{fig-2} for the $\gamma_q$ and $\gamma_s$ in NEQ and BEC, while in EQ $\gamma_q=\gamma_s=1$.}
\label{fig-3}
\end{figure*}

 The $\pi^0$ mesons will condense first, because they are the lightest particles. The $\pi^0$ multiplicity is not measured in Pb+Pb collisions at the LHC yet. Therefore we add
 the estimate for the number of $\pi^0$ mesons as $\pi^0=(\pi^++\pi^-)/2$ and fit it together with all other available particle multiplicities. The results are shown in Figs.~\ref{fig-2}~ and~\ref{fig-3}. We checked that the measured $\pi^0$ spectrum~\cite{Abelev:2014ypa} agrees with our estimate. The data exist only for the range $p_T\gtrsim 700$~MeV.
 It gives just about $1/3$ of the total expected $\pi^0$ multiplicity. Therefore the measurement of the low $p_T$ spectrum of neutral pions is crucially important to judge about the Bose condensation.

One can see that the BEC and NEQ volumes coincide within the
errors, while the EQ volume is substantially larger. This is in
agreement with the calculations of other
authors~\cite{Petran:2013qla,Petran:2013lja,Floris:2014pta} in the
EQ and NEQ models. The temperature in EQ is almost constant and is
between $150-160$~MeV, as is expected for the equilibrium.
On the other hand the BEC temperature demonstrates an interesting
centrality dependance. In most central collisions it is close to
the temperature in NEQ, while at very peripheral collisions it
approaches the EQ temperature.
%
%Such a behavior is possible if pion condensation is stronger at small centrality.
The $\gamma_q$ and $\gamma_s$ parameters also strongly depend on
centrality, see Fig.~\ref{fig-3}.
At small centralities the $\gamma_q$ and $\gamma_s$ values in BEC
are close to those in NEQ, while at high centrality both
$\gamma_q$ and $\gamma_s$ approach unity. The $\gamma$'s in BEC
are always smaller than in NEQ. It means that the inclusion of the
ground state decreases the chemical potential and the number of
particles in the condensate. However, the detailed determination
of the condensate rate as a function of centrality requires a
separate study~\cite{Begun:2015ifa}.
%
%It could help to describe the heavy strange particles, as noted in~\cite{Begun:2014rsa}.
%However, the error bars are too big to start the cumbersome analysis of the spectra with the new BEC parameters.
%At first, one has to find a way to make more precise calculations~\cite{Begun&Flork}.

%%%%%%%%%%%%%%%%%%%%%%%%%%%%%%%%%%%%%%%%%%%%%%%%%%%%%%%%%%
%
\section{Conclusions} \label{sec-3}
%
%%%%%%%%%%%%%%%%%%%%%%%%%%%%%%%%%%%%%%%%%%%%%%%%%%%%%%%%%%
%
 The non-equilibrium thermal model combined with the single freeze-out scenario explains very well the spectra of light particles.
 It eliminates the proton anomaly and explains the low-$p_T$ enhancement of pions.
 This enhancement may be interpreted as a signature of the onset of pion condensation in heavy-ion collisions at the LHC.
 Since the difference between equilibrium and non-equilibrium models strongly increases at low $p_T$, it would be interesting to see the measurements of the charged pion spectrum at smaller values
 of $p_T$ than those available at the moment. The same is even more important for the $\pi^0$ meson spectrum, because neutral pions condense first.
\vspace{1cm}
%%%%%%%%%%%%%%%%%%%%%%%%%%%%%%%%%%%%%%%%%%%%%%%%%%%%%%%%%

%%%%%%%%%%%%%%%%%%%%%%%%%%%%%%%%%%%%%%%%%%%%%%%%%%%%%%%%%%%%%%%%%%%%%%%%%%%%%%%%%%%%%%%%%%%%%%%%%%%
{\bf Acknowledgments:}
%%%%%%%%%%%%%%%%%%%%%%%%%%%%%%%%%%%%%%%%%%%%%%%%%%%%%%%%%%%%%%%%%%%%%%%%%%%%%%%%%%%%%%%%%%%%%%%%%%%
\vspace{0.5cm}

I would like to thank Wojtek Florkowski for fruitful discussions
and advices. This work was supported by Polish National Science
Center grant No. DEC-2012/06/A/ST2/00390.

%
% BibTeX or Biber users please use (the style is already called in the class, ensure that the "woc.bst" style is in your local directory)
% \bibliography{name or your bibliography database}

\begin{thebibliography}{100}
%

\bibitem{Cleymans:1992zc}
  J.~Cleymans and H.~Satz,
  %``Thermal hadron production in high-energy heavy ion collisions,''
  Z.\ Phys.\ C {\bf 57}, 135 (1993)
  [hep-ph/9207204].

\bibitem{Cleymans:1999st}
  J.~Cleymans and K.~Redlich,
  %``Chemical and thermal freezeout parameters from 1-A/GeV to 200-A/GeV,''
  Phys.\ Rev.\ C {\bf 60}, 054908 (1999)
  [nucl-th/9903063].

\bibitem{BraunMunzinger:2003zd}
  P.~Braun-Munzinger, K.~Redlich and J.~Stachel,
  %``Particle production in heavy ion collisions,''
  In R.~C.~Hwa, X.~N.~Wang \textit{Quark gluon plasma} World Scientific Publishing, 2004, 491-599
  [nucl-th/0304013].

\bibitem{Becattini:2003wp}
  F.~Becattini, M.~Gazdzicki, A.~Keranen, J.~Manninen and R.~Stock,
  %``Chemical equilibrium in nucleus nucleus collisions at relativistic energies,''
  Phys.\ Rev.\ C {\bf 69}, 024905 (2004)
  [hep-ph/0310049].

\bibitem{Becattini:1995if}
  F.~Becattini,
  %``A Thermodynamical approach to hadron production in e+ e- collisions,''
  Z.\ Phys.\ C {\bf 69}, 485 (1996).

\bibitem{Becattini:1997rv}
  F.~Becattini and U.~W.~Heinz,
  %``Thermal hadron production in p p and p anti-p collisions,''
  Z.\ Phys.\ C {\bf 76}, 269 (1997)
  [Erratum-ibid.\ C {\bf 76}, 578 (1997)]
  [hep-ph/9702274].

\bibitem{Begun:2013nga}
  V.~Begun, W.~Florkowski and M.~Rybczynski,
  %``Explanation of hadron transverse-momentum spectra in heavy-ion collisions at $\sqrt s_{NN} =$ 2.76 TeV within chemical non-equilibrium statistical hadronization model,''
  Phys.\ Rev.\ C {\bf 90}, no. 1, 014906 (2014)
  [arXiv:1312.1487 [nucl-th]].

\bibitem{Begun:2014rsa}
  V.~Begun, W.~Florkowski and M.~Rybczynski,
  %``Transverse-momentum spectra of strange particles produced in Pb+Pb collisions at $\sqrt{s_{\rm NN}}=2.76$ TeV in the chemical non-equilibrium model,''
  Phys.\ Rev.\ C {\bf 90}, no. 5, 054912 (2014)
  [arXiv:1405.7252 [hep-ph]].

\bibitem{Stachel:2013zma}
  J.~Stachel, A.~Andronic, P.~Braun-Munzinger and K.~Redlich,
  %  %``Confronting LHC data with the statistical hadronization model,''
  J.\ Phys.\ Conf.\ Ser.\  {\bf 509}, 012019 (2014)
  [arXiv:1311.4662 [nucl-th]].

\bibitem{Becattini:2012xb}
  F.~Becattini, M.~Bleicher, T.~Kollegger, T.~Schuster, J.~Steinheimer and R.~Stock,
  %``Hadron Formation in Relativistic Nuclear Collisions and the QCD Phase Diagram,''
  Phys.\ Rev.\ Lett.\  {\bf 111}, 082302 (2013)
  [arXiv:1212.2431 [nucl-th]].

\bibitem{Noronha-Hostler:2014usa}
  J.~Noronha-Hostler and C.~Greiner,
  %``Suppression of the LHC $p/\pi$ ratio due to the QCD mass spectrum,''
  arXiv:1405.7298 [nucl-th].

\bibitem{Noronha-Hostler:2014aia}
  J.~Noronha-Hostler and C.~Greiner,
  %``Understanding the $p/\pi$ ratio at LHC due to QCD mass spectrum,''
  arXiv:1408.0761 [nucl-th].

\bibitem{Chatterjee:2013yga}
  S.~Chatterjee, R.~M.~Godbole and S.~Gupta,
  %``Strange freezeout,''
  Phys.\ Lett.\ B {\bf 727}, 554 (2013)
  [arXiv:1306.2006 [nucl-th]].

\bibitem{Petran:2013qla}
  M.~Petran and J.~Rafelski,
  %``Universal hadronization condition in heavy ion collisions at $\sqrt{s_\mathrm{NN}}= 62$ GeV and at $\sqrt{s_\mathrm{NN}}=2.76$ TeV,''
  Phys.\ Rev.\ C {\bf 88}, no. 2, 021901 (2013)
  [arXiv:1303.0913 [hep-ph]].

\bibitem{Petran:2013lja}
  M.~Petran, J.~Letessier, V.~Petracek and J.~Rafelski,
  %``Hadron production and quark-gluon plasma hadronization in Pb-Pb collisions at $\sqrt{s_{NN}}=2.76$ TeV,''
  Phys.\ Rev.\ C {\bf 88}, no. 3, 034907 (2013)
  [arXiv:1303.2098 [hep-ph]].

%\bibitem{Melo&Tomasik}
% I.~Melo, B.~Tomasik, poster presentation at the {\it Quark Matter 2014} conference, Darmstadt, Germany, 2014.
%%``Freeze-out state from analysis of transverse momentum spectra in Pb+Pb collisions at 2.76 ATeV,''

\bibitem{Floris:2014pta}
  M.~Floris,
  %``Hadron yields and the phase diagram of strongly interacting matter,''
  Nucl.\ Phys.\ A {\bf 931}, 103 (2014)
  [arXiv:1408.6403 [nucl-ex]].

\bibitem{Broniowski:2001we}
  W.~Broniowski and W.~Florkowski,
  %``Explanation of the RHIC p(T) spectra in a thermal model with expansion,''
  Phys.\ Rev.\ Lett.\  {\bf 87}, 272302 (2001)
  [nucl-th/0106050].

\bibitem{Broniowski:2001uk}
  W.~Broniowski and W.~Florkowski,
  %``Strange particle production at RHIC in a single freezeout model,''
  Phys.\ Rev.\ C {\bf 65}, 064905 (2002)
  [nucl-th/0112043].

\bibitem{Rybczynski:2012ed}
  M.~Rybczynski, W.~Florkowski and W.~Broniowski,
  %``Single-freeze-out model for ultra relativistic heavy-ion collisions at $\sqrt{s_{\rm NN}}=2.76$ TeV and the LHC proton puzzle,''
  Phys.\ Rev.\ C {\bf 85}, 054907 (2012)
  [arXiv:1202.5639 [nucl-th]].

\bibitem{Abelev:2012wca}
  B.~Abelev {\it et al.}  [ALICE Collaboration],
  %``Pion, Kaon, and Proton Production in Central Pb--Pb Collisions at $\sqrt{s_{NN}} = 2.76$ TeV,''
  Phys.\ Rev.\ Lett.\  {\bf 109}, 252301 (2012)
  [arXiv:1208.1974 [hep-ex]].

\bibitem{Abelev:2013vea}
  B.~Abelev {\it et al.}  [ALICE Collaboration],
  %``Centrality dependence of $\pi$, K, p production in Pb-Pb collisions at $\sqrt{s_{NN}}$ = 2.76 TeV,''
  Phys.\ Rev.\ C {\bf 88}, 044910 (2013)
  [arXiv:1303.0737 [hep-ex]]

\bibitem{Gale:2012rq}
  C.~Gale, S.~Jeon, B.~Schenke, P.~Tribedy and R.~Venugopalan,
  %``Event-by-event anisotropic flow in heavy-ion collisions from combined Yang-Mills and viscous fluid dynamics,''
  Phys.\ Rev.\ Lett.\  {\bf 110}, 012302 (2013)
  [arXiv:1209.6330 [nucl-th]].

\bibitem{Molnar:2014zha}
  E.~Molnar, H.~Holopainen, P.~Huovinen and H.~Niemi,
  %``Influence of temperature-dependent shear viscosity on elliptic flow at backward and forward rapidities in ultrarelativistic heavy-ion collisions,''
  Phys.\ Rev.\ C {\bf 90}, no. 4, 044904 (2014)
  [arXiv:1407.8152 [nucl-th]].

\bibitem{vanderSchee:2013pia}
  W.~van der Schee, P.~Romatschke and S.~Pratt,
  %``Fully Dynamical Simulation of Central Nuclear Collisions,''
  Phys.\ Rev.\ Lett.\  {\bf 111}, no. 22, 222302 (2013)
  [arXiv:1307.2539].

\bibitem{Kisiel:2005hn}
  A.~Kisiel, T.~Taluc, W.~Broniowski and W.~Florkowski,
  %``THERMINATOR: THERMal heavy-IoN generATOR,''
  Comput.\ Phys.\ Commun.\  {\bf 174}, 669 (2006)
  [nucl-th/0504047].

\bibitem{Chojnacki:2011hb}
  M.~Chojnacki, A.~Kisiel, W.~Florkowski and W.~Broniowski,
  %``THERMINATOR 2: THERMal heavy IoN generATOR 2,''
  Comput.\ Phys.\ Commun.\  {\bf 183}, 746 (2012)
  [arXiv:1102.0273 [nucl-th]].

\bibitem{Blaizot:2011xf}
  J.~-P.~Blaizot, F.~Gelis, J.~-F.~Liao, L.~McLerran and R.~Venugopalan,
  %``Bose--Einstein Condensation and Thermalization of the Quark Gluon Plasma,''
  Nucl.\ Phys.\ A {\bf 873}, 68 (2012)
  [arXiv:1107.5296 [hep-ph]].

 \bibitem{Blaizot:2013lga}
  J.~-P.~Blaizot, J.~Liao and L.~McLerran,
  %``Gluon Transport Equation in the Small Angle Approximation and the Onset of Bose-Einstein Condensation,''
  Nucl.\ Phys.\ A {\bf 920}, 58 (2013)
  [arXiv:1305.2119 [hep-ph]].

\bibitem{Gelis:2014tda}
  F.~Gelis,
  %``Initial state in relativistic nuclear collisions and Color Glass Condensate,''
  Nucl.\ Phys.\ A {\bf 931}, 73 (2014)
  [arXiv:1412.0471 [hep-ph]].

\bibitem{Begun:2008hq}
  V.~V.~Begun and M.~I.~Gorenstein,
  %``Bose-Einstein Condensation in the Relativistic Pion Gas: Thermodynamic Limit and Finite Size Effects,''
  Phys.\ Rev.\ C {\bf 77}, 064903 (2008)
  [arXiv:0802.3349 [hep-ph]].

\bibitem{Petran:2013dva}
  M.~Petran, J.~Letessier, J.~Rafelski and G.~Torrieri,
  %``SHARE with CHARM,''
  Comput.\ Phys.\ Commun.\  {\bf 185}, 2056 (2014)
  [arXiv:1310.5108 [hep-ph]].

\bibitem{Abelev:2014ypa}
  B.~B.~Abelev {\it et al.}  [ALICE Collaboration],
  %``Neutral pion production at midrapidity in pp and Pb-Pb collisions at $\sqrt{s_{{\mathrm {NN}}}}= 2.76\,{\mathrm {TeV}}$,''
  Eur.\ Phys.\ J.\ C {\bf 74}, no. 10, 3108 (2014)
  [arXiv:1405.3794 [nucl-ex]].

\bibitem{Begun:2015ifa}
  V.~Begun and W.~Florkowski,
  %``Bose-Einstein condensation of pions in heavy-ion collisions,''
  arXiv:1503.04040 [nucl-th].

\end{thebibliography}
%
% Non-BibTeX users please use
%

\end{document}